\documentclass[twocolumn,amsmath,amssymb,aps,prl,reprint,longbibliography]{revtex4-1}
\usepackage{graphicx}
\usepackage{xcolor}
\begin{document}

\title{Active Phases for Particles on Resource Landscapes}
 
\author{L. Varga$^1$, A. Lib{\' a}l$^1$, C. J. O. Reichhardt$^2$, and C. Reichhardt$^2$}
\affiliation
{$^1$ Mathematics and Computer Science Department, Babe{\c s}-Bolyai University,
Cluj 400084, Romania\\
$^2$ Theoretical Division and Center for Nonlinear Studies,
Los Alamos National Laboratory, Los Alamos, New Mexico 87545, USA}

\date{\today}

\begin{abstract}
We introduce an active matter model composed of sterically interacting particles which absorb resources from a substrate and move in response to resource gradients. For varied ratios of absorption rate to substrate recovery rate, we find a variety of phases including periodic waves, partial clustering, stochastic motion, and a frozen state. If passive particles are added, they can form crystalline clusters in an active fluid. This model could be implemented using colloidal systems on feedback landscapes and can provide a soft matter realization of excitable media and ecological systems.   

\end{abstract}

\maketitle

Active matter exhibits self-mobility \cite{Marchetti13,Cates15,Bechinger16},
which can appear in biological \cite{Berg83,DellArciprete18},
social \cite{Helbing01,Couzin05}, robotic \cite{Deblais18,Wang21}, 
and soft matter systems \cite{Palacci13,Buttinoni13}. 
For active matter composed of particles, the motility can be modeled as  
a motor force providing run-and-tumble or driven diffusive propulsion
\cite{Marchetti13,Cates15,Bechinger16}, and additional 
dynamics can be included which 
induce different types of flocking behaviors \cite{Vicsek12,Barberis16}. 
Active particle assemblies exhibit a variety of phenomena that are absent 
in Brownian systems, such as the motility-induced
phase separation or clustering that
can arise even when all the pairwise particle-particle interactions
are repulsive \cite{Cates15,Palacci13,Buttinoni13,Fily12,Redner13}.  
In certain biological active matter systems,
the motion can be 
affected by local or global gradients  
in the environment, producing effects such as
chemotaxis
or drift of particles in certain directions \cite{Berg83,Stark18}.
There are also numerous examples of
active particles that are coupled to
complex environments \cite{Cates15}, leading to phenomena such 
as
active matter 
ratchets \cite{Galajda07,Reichhardt17a},
trapping or substrate-induced clustering
\cite{Cates15,Reichhardt14,Morin17,Bhattacharjee19a},
or topotaxis, where spatial gradients in the landscape
generate directed motion \cite{Schakenraad20,Vergassola07}.

It is also possible for activity to arise not
from a motor force carried by each particle,
but instead
from forces induced by a substrate. 
Such situations can arise in
time dependent environments \cite{Couzin05,Cenzer19},
particles coupled to excitable media \cite{Zykov18},  
and colloids on feedback substrates \cite{Bauerle18}.
Recently, Wang {\it et al.} \cite{Wang21} introduced an ecology-inspired
active matter system of robots 
interacting with a resource substrate where the robots consume the resources
and are attracted to regions with
the highest resource concentration.   
This system exhibited numerous phases such as crystalline, liquid, glass,
and jammed states.

Here we propose a new type of soft matter active system
for an assembly of sterically repulsive particles 
that couple to a resource landscape.
The particle motion is governed by local
gradients of the resource concentration. 
When a single particle sits over a
group of resource sites,
it experiences a net force
directed
towards the sites containing the highest resource levels.
Collisions between adjacent particles cause localized pinning-depinning
transitions that make the behavior distinct from
that found in a reaction-diffusion system.
Resources are depleted from sites occupied by the particle at a fixed rate,
while all sites in the system recover
resource levels at a different
fixed rate up to a maximum resource value.

We find that for high absorption rates and low recovery rates,
particle motion occurs in periodic bursts of 
activity via the propagation of waves through the system,
similar to the reaction diffusion patterns
found in excitable media \cite{Zykov18,Zaikin70,Sakurai02}. 
As the recovery rate is increased, the pulse frequency increases 
until a transition occurs to a continuously fluctuating or chaotic fluid state. 
When the recovery rate is larger than the absorption rate, 
the particles become frozen into place since a resource gradient never
develops.
For some rate combinations, the particles form a partially 
clustered phase due to collisions of oppositely propagating waves.
When we add a second species of passive particles
that do not couple to the substrate but only interact sterically with
the active particles,
we find
a crystallization of the passive particles
within the fluctuating fluid phase
similar to the motility induced phase separation observed for
active particles with motor forces.

Our system could be realized
using colloids interacting with an optical substrate
in the presence of feedback mechanisms
\cite{Bauerle18,Lavergne19} or colloids
coupled to an excitable medium
where the presence of a colloid
locally alters the diffusion rate in the medium. 
Our results indicate that soft active matter could be used to explore
excitable media and ecologically inspired systems. 

{\it Simulation and System---} 
We consider a two-dimensional system of size $L_x \times L_y$ with $L_x=100$
and $L_y=200$  
containing $N_p=3500$
particles of diameter $d = 1.0$ with steric repulsive interactions.  
The particle area coverage is $\phi=\pi d^2/(4L_xL_y)=0.549$,
well below the jamming density of $\phi_J=0.9$. 
The particles also couple to a substrate consisting
of a fine mesh of $N_g=80000$
grid sites, each of size $l_g \times l_g$ with $l_g=0.5$.
A given grid site interacts with at most one particle at a time.
Defining $g^i_c(t)$ to be the grid site closest to the center of particle $i$
at time $t$, the ``occluded sites'' which exert an attractive force
on the particle are defined
to be the eight nearest neighbors of
$g^i_c(t)$ (including diagonal nearest neighbors)
and the four linear next-nearest neighbors of $g^i_c(t)$.
The overdamped equation of motion of particle $i$ is given by
\begin{equation}  
\eta {\bf v}_{i} = {\bf F}_i^{\rm tot}={\bf F}^{pp}_{i} + {\bf F}^{\rm g}_i \ ,
\end{equation}
where  ${\bf v}_i=d{\bf r}_i/dt$ is the velocity
of particle $i$ and $\eta$
is the damping constant which is set to unity. 
The particle-particle interaction force
${\bf F}_i^{pp} = \sum^{N_p}_{j=1}k(d-r_{ij})\Theta(d-r_{ij}){\bf \hat{r}}_{ij}$
arises from harmonic repulsion with spring constant $k=20$,
$r_{ij}=|{\bf r}_i-{\bf r}_j|$, and ${\bf \hat{r}}_{ij}=({\bf r}_i-{\bf r}_j)/r_{ij}$, where $\Theta$ is the Heaviside step function.
The attractive substrate force is summed over the 12 grid sites
occluded by particle $i$,
${\bf F}^{g}_i=-\sum_{k=1}^{12} S_{g}^k{\bf \hat{r}}_{ik}$,
where ${\bf \hat{r}}_{ik}=({\bf r}_i - {\bf r}_k)/r_{ik}$,
$r_{ik}=|{\bf r}_i-{\bf r}_k|$,
and
${\bf r}_k$ is the center of grid site $k$.
The
attraction is
proportional to the resource level $S_{g}^k(t)$ available at grid site $k$
at time t,
which ranges from the completely depleted value $S^k_{g}=0$
to the saturated maximum value
$S^k_{g}=1$.
If all sites occluded by a particle have the same value of $S_g$, the
attractive forces cancel and the particle does not move.
The evolution of the resource level at grid site $l$ is
$S_{g}^l(t+\Delta t)=S_{g}^l(t)-O_g^lr_{\rm abs}+r_{\rm rec}$, where
$r_{\rm abs}$ is the rate at which a particle absorbs resources from
the grid site and $O_g^l=1$ (0) if the grid site is occluded
(unoccupied).
All grid sites recover resources at rate $r_{\rm rec}$,
but the value of $S_g^l$ is not allowed
to exceed $S_g^l=1$.
For initialization,
$S_g^l(0)$ for each site is set randomly to values between 0 and 1, while
the particles are placed at randomly chosen locations with
particle overlap forbidden.
To characterize the behavior, we measure
the fraction of particles that are moving at each instant,
$M=N_p^{-1}\sum_{i}^{N_p} \Theta(|{\bf F}_i^{\rm tot}|-0.01)$,
its time average $\langle M\rangle$, and its
standard deviation $\sigma_M$. We also measure $C_l$, the fraction of particles
belonging to the largest cluster in the system,
where particles that are in contact with each other
are defined as belong to a given
cluster.

\begin{figure}
  \includegraphics[width=3.0in]{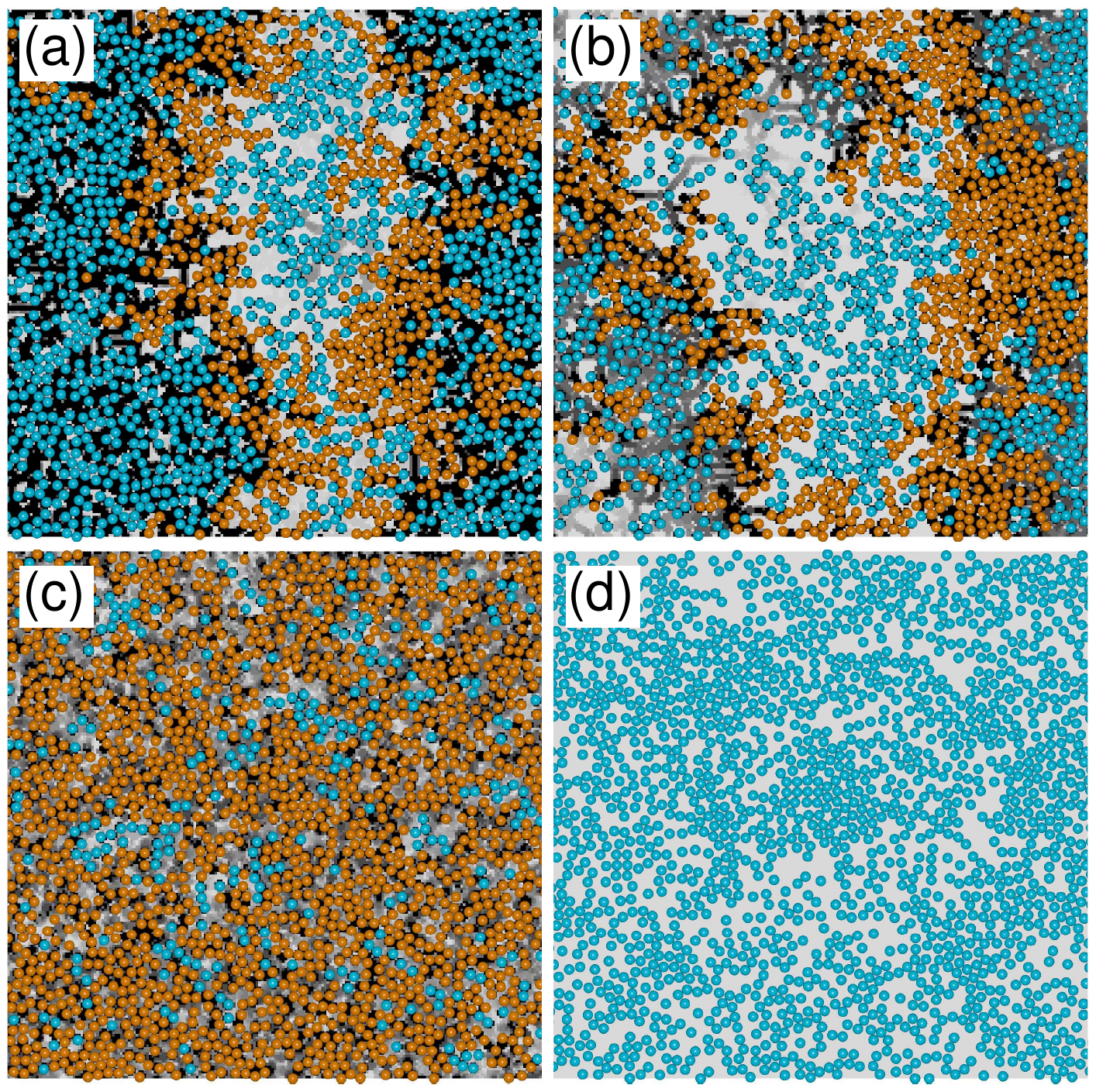}
  \caption{Simulation snapshots
of a $100 \times 100$ portion of the sample
showing moving (orange) and stationary (blue)
particles on a resource field with resource values ranging   
from $S_g=0$ (black) to $S_g=1$ (white).
(a)
Particles move in propagating waves
for high absorption rate
$r_{\rm abs}=8.5 \times 10^{-3}$
and low recovery rate 
$r_{\rm rec} = 5 \times 10^{-5}$.
(b)
For higher recovery rate
$r_{\rm rec}= 5 \times 10^{-4}$ at $r_{\rm abs}=8.5 \times 10^{-3}$,
pulses of motion appear more frequently and partial clustering of the
moving particles occurs.
(c)
When $r_{\rm rec}$ is comparable to $r_{\rm abs}$, 
a continuously moving or fluid phase appears,
as shown for
$r_{\rm abs} = 2.5 \times 10^{-3}$ and $r_{\rm rec} = 7 \times 10^{-4}$.
(d) The system is frozen when
$r_{\rm rec}>r_{\rm abs}$,
as shown for
$r_{\rm abs} = 4 \times 10^{-4}$ and $r_{\rm rec} = 8 \times 10^{-4}$. 
Movies of these states appear in the supplemental material \cite{Supplemental}.
}
\label{fig:1}
\end{figure}

{\it Results---}
In Fig.~\ref{fig:1} we show snapshots of a portion of the
system in different dynamic phases.   
At high absorption rates and low recovery
rates, illustrated in Fig.~\ref{fig:1} at
$r_{\rm abs} = 8.5\times 10^{-3}$ and $r_{\rm rec} = 5\times 10^{-5}$,
long intervals of no motion are interspersed with bursts of motion
in the form of
a front or wave which traverses the system
and leaves behind a region of depleted resource sites.
After each wave dissipates,
the system returns to a non-moving state until the grid sites have
recovered enough resources to reactivate the particles.
If we increase $r_{\rm rec}$ while holding $r_{\rm abs}$ fixed,
the time between pulses of motion is reduced, the propagating fronts
become wider, and we find
a partial clustering effect
as shown in Fig.~\ref{fig:1}(b) at
$r_{\rm abs} = 8.5 \times 10^{-3}$ and $r_{\rm rec} = 5 \times 10^{-4}$.
As the recovery rate is further increased,
the pulse frequency increases until a transition
occurs to a state in which most particles are moving most of the time,
creating a fluctuating fluid as illustrated in Fig.~\ref{fig:1}(c) for
$r_{\rm abs} = 2.5 \times 10^{-3}$ and
$r_{\rm rec} = 7 \times 10^{-4}$.
For high recovery and low absorption rates,
the system becomes trapped in a frozen phase, 
as shown in Fig.~\ref{fig:1}(d) at
$r_{\rm abs} = 4 \times 10^{-4}$ and
$r_{\rm rec} = 8 \times 10^{-4}$.
The overall behavior can be described
in ecological terminology by allowing
the resources to represent food. Oscillations occur
when there is not enough food to support continuous motion;
instead,
a food accumulation period
is required to induce movement.
Once the food has recovered, at least one
particle moves and collides with another particle,
triggering the wave of motion.
After the wave has propagated through the system,
the food is depleted again and motion does not resume until
the food has been replenished.
In the fluctuating state of Fig.~\ref{fig:1}(c), when food is depleted
at a local grid site, the particle can immediately find sufficient food
at neighboring grid sites, permitting the particle to maintain
nearly constant motion.
When the food is extremely plentiful,
a particle always finds more than enough food at its current
position, so it does not need to move.

\begin{figure}
\includegraphics[width=3.5in]{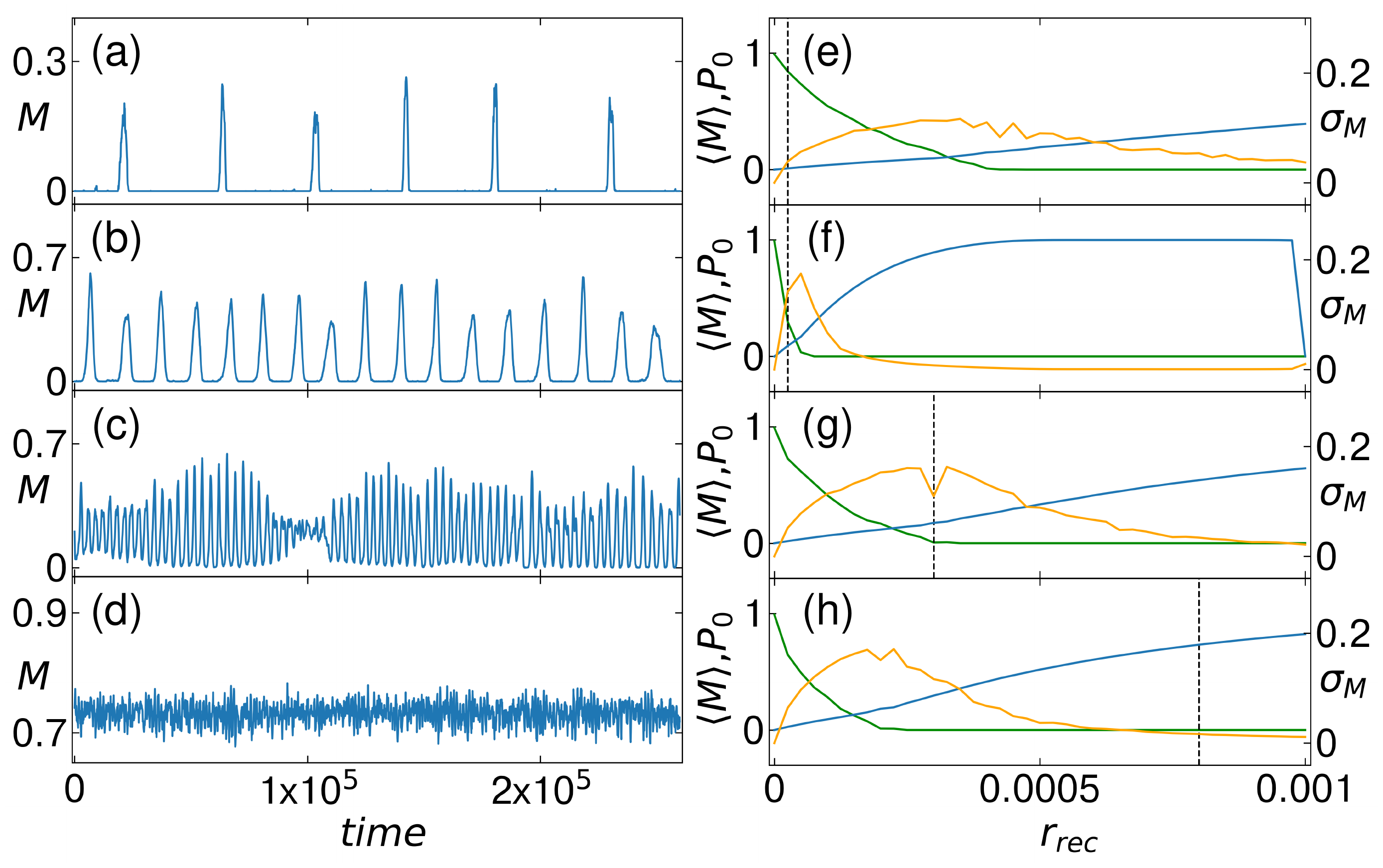}
\caption{Left column:
The fraction of moving particles $M$ versus
time in simulation steps.
(a) In phase I at
$r_{\rm rec} = 2.5 \times 10^{-5}$ and $r_{\rm abs} = 0.01$,
there are slow periodic oscillations in $M$.
(b) At
$r_{\rm rec} = 2.5 \times 10^{-5}$ and $r_{\rm abs} = 0.001$,
we find phase I motion but with shorter intervals between bursts. 
(c)
At $r_{\rm rec} = 3 \times 10^{-4}$ and $r_{\rm abs} = 0.006$,
the system is in phase II with rapid periodic motion.
(d) For
$r_{\rm rec} = 8 \times 10^{-4}$ and $r_{\rm abs} = 0.004$,
we find
phase III continuous motion.
Right column: The time averaged fraction of moving
particles $\langle M\rangle$ (blue),
its standard deviation $\sigma_M$ (orange),
and the fraction of time $P_0$ during which the motion is zero
(green) versus
recovery rate $r_{\rm rec}$
at absorption rates of $r_{\rm abs}=$
(e) 0.01, (f) 0.001, (g) $0.006$, and (g) $0.004$ matching the
rates shown in panels (a) through (d). Dashed lines in the left column
indicate the value of $r_{\rm rec}$ illustrated in the right column.
}
\label{fig:2}
\end{figure}

In Fig.~\ref{fig:2}(a) we plot the fraction of moving particles
$M$ versus time for a system with propagating waves
at
$r_{\rm rec} = 2.5 \times 10^{-5}$ and $r_{\rm abs} = 0.01$,
where long intervals of no activity are
interspersed with sharp bursts of motion.
For the same system with a lower $r_{\rm abs}=0.001$
in Fig.~\ref{fig:2}(b),
$M$ still has windows of close to zero activity, but
the interval between activity bursts is almost 3.5 times shorter.
In Fig.~\ref{fig:2}(c),
where $r_{\rm rec}=3 \times 10^{-4}$ and $r_{\rm abs}=0.006$,
we find
a more rapid oscillating behavior associated with the propagation
of multiple fronts
though the system with slightly different periods,
producing beat-like patterns in $M$.
For
$r_{\rm rec} = 8 \times 10^{-4}$ and $r_{\rm abs} = 0.004$ 
in Fig.~\ref{fig:2}(d), 
the motion is continuous and chaotic, similar to what is shown
in Fig.~\ref{fig:1}(c).
In the frozen phase, such as that illustrated in
Fig.~\ref{fig:1}(d),
we find $M=0$
apart from a brief initial transient.

In Fig.~\ref{fig:2}(e-h) we plot the time averaged fraction of moving
particles
$\langle M\rangle$, its standard deviation $\sigma_M$, and the fraction of
time $P_0$ during which $M=0$
versus $r_{\rm rec}$ at
$r_{\rm abs}=0.01$, 0.001, 0.006, and $0.004$, respectively.
We use these measures to delineate the dynamic phases of the system.
In phase I, intermittent or periodic burst
behavior of the type shown in Figs.~\ref{fig:2}(a,b)
appears.
Here, a single moving front of
particles activates adjacent particles by pushing them
forward, creating a wave-like excitation.
The motion depletes the resources of the substrate,
so the particles become motionless after the wave passes,
and a finite recovery time is necessary before the substrate can
support particle motion again.
A long recovery time appears as large values of $P_0$ for low $r_{\rm rec}$.
Throughout phase I,
$\langle M\rangle$ increases with increasing $r_{\rm rec}$, but it
exhibits large variations as indicated by the increase
in $\sigma_M$ with increasing $r_{\rm rec}$.
Phase II is defined to occur when $P_0<0.5$ but
the motion is still periodic.
In this regime
the waiting time for the substrate recovery is reduced,
and motion
initiates at
multiple points, producing
multiple simultaneously propagating fronts.
The result is very rapid oscillations
in $M$ and the appearance of a
peak in $\sigma_M$.
At even higher $r_{\rm rec}$, there is a transition to phase III,
consisting
of continuous or chaotic motion.
Here, $\sigma_M$
decreases
with increasing $r_{\rm rec}$
as the intermittent nature of the motion disappears,
while $P_0$ drops and $\langle M\rangle$ becomes large.
When $r_{\rm rec}$ is increased further, there is a sudden transition to
the frozen state, termed phase IV, where the recovery is sufficiently rapid
that all grid sites remain saturated to $S_g=1$ and gradients in $S_g$,
required to produce motion of the particles, do not develop.
Figure~\ref{fig:2}(f) shows that in a sample with
$r_{\rm abs} = 0.001$,
the frozen phase IV appears
at $r_{\rm rec} = 0.001$, where $\langle M\rangle$ drops abruptly to zero.
The oscillating motion
in phases I and II and the transition to the chaotic phase III motion
are similar to what is observed
in reaction-diffusion systems \cite{Zykov18,Zaikin70,Sakurai02}. 
In our system, the fronts are correlated with translating
motion of the particles rather than with reaction activity.

\begin{figure}
\includegraphics[width=3.5in]{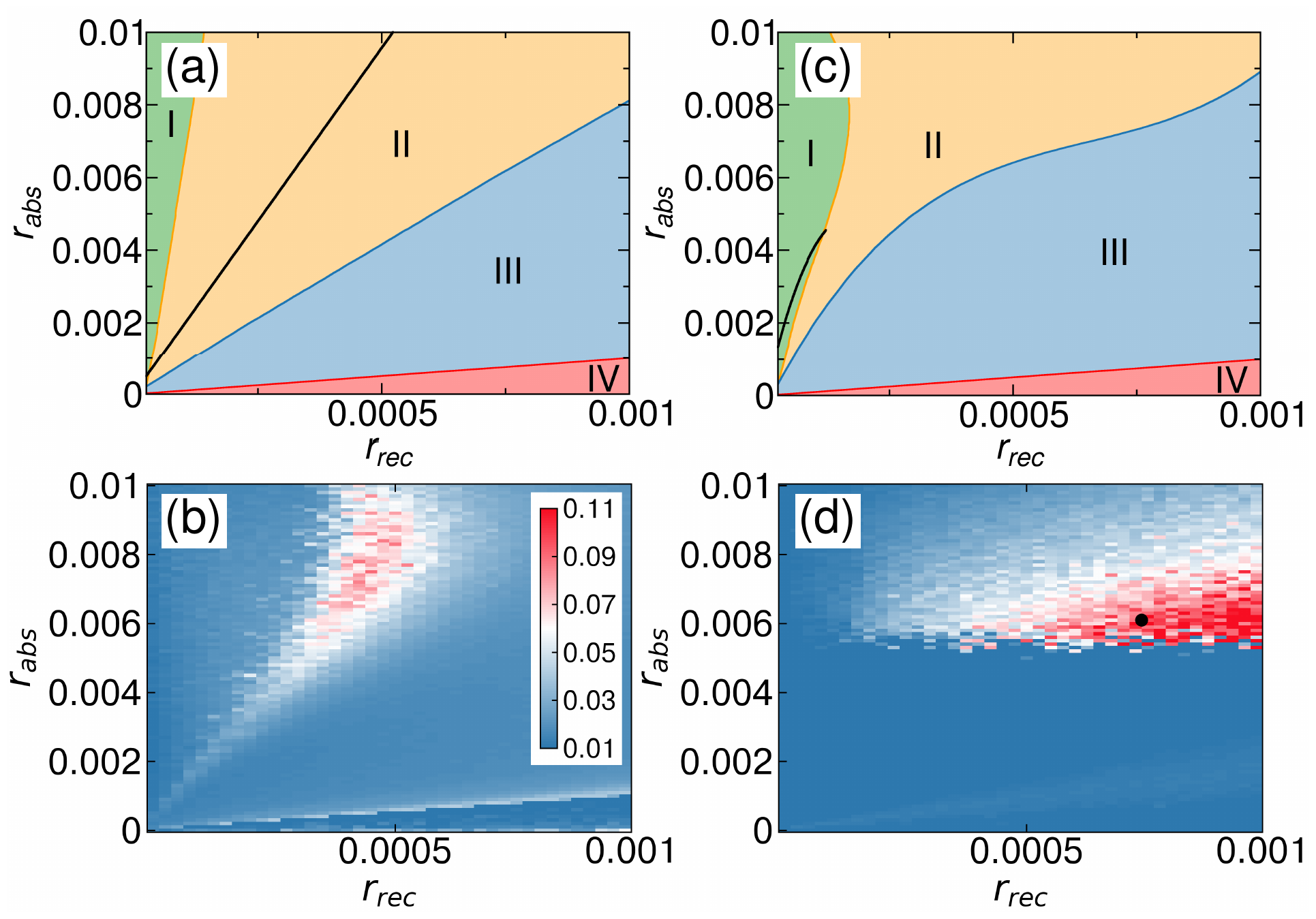}
\caption{(a)
Dynamic phase diagram as a function of $r_{\rm abs}$ versus $r_{\rm rec}$
showing phases I (periodic bursts, green), II (oscillations, orange),
III (continuous motion, purple), and IV (frozen, red).
The black line indicates the location of the peak in $\sigma_M$.
(b) Heat map of the fraction $C_l$ of particles in
the largest cluster for the same system.
Cluster size is maximized in phase II near the location of the
peak in $\sigma_M$.
(c) Dynamic phase diagram as a function of $r_{\rm abs}$ versus $r_{\rm rec}$
in a system where
half of the particles are passive and do not interact with the substrate.
(d) The corresponding heat map of $C_l$
showing that crystallization now occurs in phase III,
where the passive particles form
triangular crystallites as illustrated in Fig.~\ref{fig:4}.  
}
\label{fig:3}
\end{figure}

By conducting a series of simulations in which we
measure $\langle M\rangle$ and $P_0$,
we construct a
dynamic phase diagram as a function of $r_{\rm abs}$ versus $r_{\rm rec}$ 
as shown in Fig.~\ref{fig:3}(a).
The transition from I to II appears at $r_{\rm abs} = 75 r_{\rm rec}$,       
the II-III crossover occurs
when $r_{\rm abs} \approx 7.5 r_{\rm rec}$, and the
transition from III to IV
falls at $r_{\rm abs} = r_{\rm rec}$.
In Fig.~\ref{fig:3}(b), we plot a heat map of $C_l$, the fraction of particles
in the largest cluster, as a function of $r_{\rm abs}$ versus
$r_{\rm rec}$ for the same system.
The partial clustering state
appears in phase II close to the peak value of
$\sigma_M$, indicated by a black line in Fig.~\ref{fig:3}(a).
In general, clustering occurs when the
pulses of motion begin to overlap.
In phase III, the motion becomes too rapid for 
cluster formation to be possible.

\begin{figure}
\includegraphics[width=3.5in]{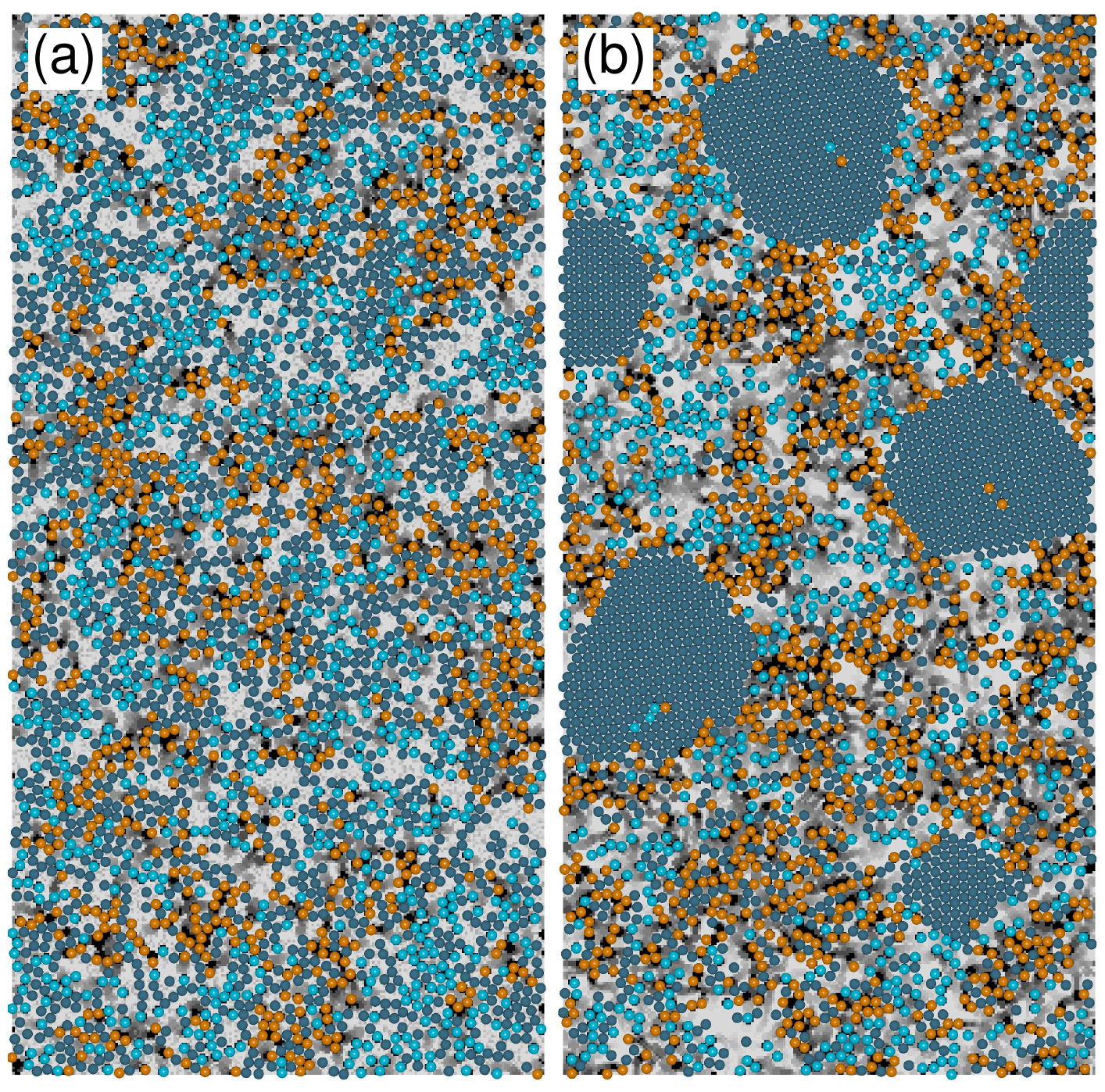}
\caption{Simulation snapshots of the entire sample showing moving (orange) and
stationary (blue) active particles along with passive (red)
particles on a resource
field with values ranging from $S_g=0$ (black) to $S_g=1$ (white) in
a sample with $r_{\rm abs} = 0.006$ and $r_{\rm rec} = 0.001$.
(a) At early times, the system forms a
phase III uniform fluctuating state.
(b) At later times, the passive particles
are pushed together into crystalline clusters.
Movies of these states appear in the supplemental material \cite{Supplemental}.
}
\label{fig:4}
\end{figure}

We have also considered a system in which half of the particles are passive
and interact only with other particles but not with the substrate.
As shown in the dynamic phase diagram of Fig.~\ref{fig:3}(c),
we find the same four phases which appeared
in the system containing only active particles;
however, the range of parameters over which phase III
extends is expanded.
In phase II there is no partial clustering; instead,
a phase separated crystallization state 
appears in region III,
as illustrated in the heat map of $C_l$ in Fig.~\ref{fig:3}(d). 
In the crystallized state,
the passive particles form dense clusters with triangular ordering.
The behavior at early and later times is illustrated
in Fig.~\ref{fig:4}(a,b)
for the system in Fig.~\ref{fig:3}(d) at
$r_{\rm abs} = 0.006$ and $r_{\rm rec} = 0.001$.
At early 
times, the particles are in a uniform fluid state,
but at later times, dense 
crystalline clusters of passive particles coexist
with a fluid of active particles,
similar to what is observed in motility induced phase separation systems
\cite{Cates15,Bechinger16,Palacci13,Buttinoni13,Redner13}.  
The crystallization occurs when $r_{\rm rec}$ is high and
$r_{\rm abs}$ is at intermediate values. 
For lower $r_{\rm abs}$, the active particles move rapidly from site to site
with a mostly Brownian characteristic, while 
for high $r_{\rm abs}$, the
active particles move only in bursts separated by long time intervals,
so there is not enough activity to generate
the phase separation.
At intermediate values of $r_{\rm abs}$, 
the substrate gradients are maximized,
generating the longest persistent intervals of active particle motion,
and producing behavior similar to that
found in run-and-tumble
or driven diffusive systems. 

{\it Summary--}
We have proposed a new type of active matter system in which particles
interact with a resource landscape
of grid sites which impart forces to the particles.
Particles are attracted  to 
sites with the most resources,
and simultaneously experience steric repulsion from other particles.
Sites underneath particles have their resources
depleted at a fixed absorption rate,
while all sites recover their resources at a different fixed
recovery rate up to a maximum value.
When the absorption rate is much larger than the recovery
rate,
we find
periodic pulses of propagating waves of translating particles.
As the recovery rate increases,
the time between pulses decreases
until there is a transition to a fluid-like state 
with continuous motion.
When the recovery rate is greater than or equal to the absorption rate,
the system becomes trapped in a frozen state.
When we introduce a second species of passive particles that do not couple to
the substrate but interact sterically with all other particles,
crystalline clusters of passive particles can form within the fluid state.
Our system could be realized
using colloidal particles interacting with an optical feedback light
substrate, reactive excitable media, or robotic assemblies
with an active substrate. This system could also be used
to create soft matter versions of ecological models and excitable media.  

\begin{acknowledgments}
We gratefully acknowledge the support of the U.S. Department of
Energy through the LANL/LDRD program for this work.
This work was supported by the US Department of Energy through
the Los Alamos National Laboratory.  Los Alamos National Laboratory is
operated by Triad National Security, LLC, for the National Nuclear Security
Administration of the U. S. Department of Energy (Contract No. 892333218NCA000001).
The work of L.V. and A.L. was supported by a grant of the Romanian Ministry of Education and Research,
CNCS - UEFISCDI, project number PN-III-P4-ID-PCE-2020-1301, within PNCDI III.
\end{acknowledgments}

\bibliography{mybib}
\end{document}